\documentclass[preprintnumbers, floatfix, preprintnumbers, letterpaper, twocolumn, superscriptaddress,nofootinbib]{revtex4}
\usepackage{graphicx}
\usepackage{microtype}
\usepackage{amsmath}
\usepackage{amssymb}
\usepackage{subfigure}
\usepackage{hyperref}
\usepackage{url}
\usepackage{xcolor}
\usepackage{color}
\usepackage{mathrsfs}
\usepackage{calrsfs}
\usepackage{amsfonts}
\usepackage{tabularx}
\usepackage{eucal}
\usepackage{latexsym}
\usepackage{ragged2e}
\usepackage{epsfig}
\usepackage{textcomp}
\usepackage{float}

\usepackage{caption}
\DeclareCaptionJustification{justified}{\leftskip=0pt \rightskip=0pt \parfillskip=0pt plus 1fil}
\definecolor{vividviolet}{rgb}{0.62, 0.0, 1.0}
\definecolor{amaranth}{rgb}{0.9, 0.17, 0.31}
\definecolor{palatinateblue}{rgb}{0.15, 0.23, 0.89}
\definecolor{brightpink}{rgb}{1.0, 0.0, 0.5}
\definecolor{cornflowerblue}{rgb}{0.39, 0.58, 0.93}
\definecolor{deepcarminepink}{rgb}{0.94, 0.19, 0.22}
\definecolor{radicalred}{rgb}{1.0, 0.21, 0.37}

\hypersetup{ linktoc=all,
	colorlinks, linkcolor={palatinateblue},
	citecolor={brightpink}, urlcolor={amaranth}
}

\graphicspath{{Images/}}

\def\sideremark#1{\ifvmode\leavevmode\fi\vadjust{\vbox to0pt{\vss
			\hbox to 0pt{\hskip\hsize\hskip1em
				\vbox{\hsize1.3cm\tiny\raggedright\pretolerance10000
					\noindent #1\hfill}\hss}\vbox to8pt{\vfil}\vss}}}%
\def\beq{\begin{equation}}
\def\eeq{\end{equation}}

\begin{document}
\title{Sign Switching Dark Energy from a Running Barrow Entropy}

\author{Sofia Di Gennaro}
\email{sofia.digennarox@gmail.com}
	\affiliation{Center for Gravitation and Cosmology, College of Physical Science and Technology, Yangzhou University, \\180 Siwangting Road, Yangzhou City, Jiangsu Province  225002, China}
	
\author{Yen Chin \surname{Ong}}
\email{ycong@yzu.edu.cn}
\affiliation{Center for Gravitation and Cosmology, College of Physical Science and Technology, Yangzhou University, \\180 Siwangting Road, Yangzhou City, Jiangsu Province  225002, China}
\affiliation{Shanghai Frontier Science Center for Gravitational Wave Detection, School of Aeronautics and Astronautics, Shanghai Jiao Tong University, Shanghai 200240, China}

\begin{abstract}
Barrow proposed that the area law of the entropy associated with a horizon might receive a ``fractal correction'' due to quantum gravitational effects -- in place of $S\propto A$, we have instead $S\propto A^{1+\delta/2}$, where $0\leqslant \delta \leqslant 1$ measures the deviation from the standard area law ($\delta=0$). Based on black hole thermodynamics, we argue that the Barrow entropy should run (i.e., energy scale dependent), which is reasonable given that quantum gravitational corrections are expected to be important only in high energy regime. When applied to the Friedmann equation, we demonstrate the possibility that such a running Barrow entropy index could give rise to a dynamical effective dark energy, which is asymptotically positive and vanishing, but negative at the Big Bang. Such a sign switching dark energy could help to alleviate the Hubble tension. Other cosmological implications are discussed. 
\end{abstract} 

\maketitle

\section{Introduction: Barrow Entropy and Quantum Gravity}

One of the most striking properties of gravity is that black holes -- and by extension -- apparent horizons have entropy that scales with the surface area \cite{0511051v1,0704.0793}. Specifically, the Bekenstein-Hawking entropy is
\begin{equation}
S=\frac{A}{4}\frac{k_B c^3}{G\hbar},
\end{equation}
where $k_B$, $c$ and $G$ are the Boltzmann constant, the speed of light, and Newton's gravitational constant, respectively. 

In different approaches to quantum gravity, this expression receives different forms of correction (see Sec.(\ref{2})). Recently, Barrow \cite{2004.09444} proposed that due to quantum gravity corrections black holes might exist with extremely wrinkled surfaces such that the event horizon is like, e.g., a Koch fractal surface. The entropy expression becomes
\begin{equation}{\label{Barrow}}
S=\left(\frac{A}{A_0}\right)^{1+\frac{\delta}{2}},
\end{equation}
where $0 \leqslant \delta \leqslant 1$ is the parameter that governs how ``fractalized'' the surface has become and we shall refer to it as ``Barrow entropy index'' (BEI). It is not quite clear whether we should have ${S=\frac{k_B}{4}\left(\frac{ A}{\ell_p^2}\right)^{1+\delta/2}}$ or  ${S=k_B \left(\frac{A}{4\ell_p^2}\right)^{1+\delta/2}}$. Barrow worked with $S\propto A/4\approx A$ from the beginning (before the correction) and thus this issue was not discussed. However, the constant factor $1/4$, whether it is raised to the power of $1+\delta/2$ or not, does not affect the results much except possibly when dealing with microscopic black holes, since the area would be very large anyway, so we will use the form Eq.(\ref{Barrow}), in line with most literature for easy comparison. However, for definiteness, we will set $A_0=4\ell_p^2$, where $\ell_p$ is the Planck length $\ell_p=\sqrt{G\hbar/c^3}$. Hereinafter, we will employ the Planck units so that $\hbar=G=c=k_B=1$.

Cosmological and black hole shadow constraints (assuming a fixed $\delta$) has put an upper bound on $\delta$, which is typically $\delta \lesssim O(10^{-3})$ or $O(10^{-4})$ \cite{2005.10302,2010.00986,2108.10998,2203.12010,2110.07258,2205.07787}. 

In this work, we will argue that $\delta$ has to be a function of energy scale and study the implications of Barrow entropy to cosmology (by the expansion of the universe, $\delta$ is thus a function of cosmic time). Such a possibility was also raised in \cite{2203.12010,2205.04138,2112.10159}. We expect that, as quantum gravitational effects become more pronounced near the Big Bang, $\delta \to 1$. As the universe expands and cools, $\delta \to 0$ monotonically. Incidentally, in the context of an asymptotically flat Schwarzschild black hole, this would mean that, as the black hole evaporates and its temperature increases, $\delta$ would approach unity (the effect of a running BEI on Hawking evaporation will be studied elsewhere).

The running Barrow entropy index gives rise to an additional term in the modified Friedmann equation, which we propose to be identified with an effective dynamical dark energy, which is asymptotically zero (hence small at late time). Surprisingly, the effective dark energy has a negative sign right about the Big Bang, i.e., the Universe started out as an anti-de Sitter-like spacetime. Interestingly, this may help to relax the Hubble tension -- the mismatch between the locally measured expansion rate of the universe and the inferred rate from early universe via the cosmic microwave background (CMB) \cite{1907.10625, 2103.01183}.

In addition, following a recent work of Sheykhi \cite{1}, which considered the fixed $\delta$ scenario, we shall identify the correction on the right hand side (r.h.s) of the Friedmann equation as an effective modification on the gravitational constant, instead of the matter field. This has an advantage that the resulting fixed $\delta$ cosmology satisfies the generalized second law (GSL), whereas if the matter sector is modified instead of the geometry, then GSL fails \cite{2005.08258}. The viability of the GSL in the context of interacting Barrow holographic dark energy model has also been investigated in \cite{2007.16020}.

\section{Barrow Entropy Index Is Energy Scale Dependent}\label{2}

In a recent work \cite{2204.09892}, Chen et al. studied black holes in asymptotically safe gravity, in which the gravitational constant also runs \cite{0002196} (see \cite{1107.5815,1806.10147} for cosmological studies). They showed that the area law receives a well-known logarithmic correction in quantum gravity literature: 
\begin{equation}\label{log}
S = \frac{A}{4} + C \ln(A);
\end{equation}
see \cite{0409024v3} for a review. 
This at least suggests that the reverse implication -- a modification of the Bekenstein-Hawking entropy leading to a varying effective gravitational constant (as in the BEI case) is perhaps not so surprising. 

Here we remark that although the Barrow entropy is supposed to quantify a quantum gravitational correction, it is \emph{not} the -- arguably more understood -- quantum gravity effect that gives rise to Eq.(\ref{log}). (In fact, the effects of having both corrections together were considered in \cite{neto}.)
The sign of the constant $C$ in front of the logarithmic term in Eq.$(\ref{log})$ is usually negative \cite{0409024v3}, but positive in some works like \cite{2204.09892} and \cite{0807.1232}. In \cite{0210024} Gour argued that a positive constant might be required if the area fluctuation is taken into account, whereas for a fixed area, the corrections due to the number of microstates (that describe the black hole) would give a negative constant. We have not much to add to this discussion. Instead, we would like to point out that the Barrow entropy, Eq.(\ref{Barrow}), if expanded out as a series assuming small $\delta$ (when quantum gravity correction is small), yields, up to the first order in $\delta$,
\begin{equation}\label{BEI}
S = \frac{A}{4} + \frac{A}{8} \ln\left(\frac{A}{4}\right)\delta.
\end{equation}
Note the presence of the area as a coefficient of the logarithmic correction. Thus, there is a risk that the ``correction'' term can be of the same order as the leading term.
In fact, for any fixed $\delta \ll 1$, the expression will still deviate significantly from the Bekenstein-Hawking area law for a sufficiently \emph{large} black hole, precisely when we expect quantum gravity corrections to be \emph{small}. This provides a strong argument for a running BEI -- $\delta$ should at least scale inversely proportional to $A\ln(A)$ to keep the subleading term small at $O(1)$. Such a running BEI would guarantee that quantum gravity correction is small when the black hole is large (i.e., when the energy scale is small, since the Hawking temperature is inversely proportional to the mass). Another option is to consider $\delta$ that runs like $1/A$, then the Barrow entropy would just be the same as the standard logarithmic correction (at least up to the first order in $\delta$), but in this work we take the former view, with the hope of uncovering new physics.

\section{Dark Energy From a Running Barrow Entropy Index}\label{4}

Sheykhi derived the correction to Friedmann equation due to the Barrow entropy (with fixed BEI) in \cite{1}:
\begin{equation} \label{corrected}
\left(H^2 + \frac{k}{a^2}\right)^{1-\frac{\delta}{2}} = \frac{8\pi}{3}\rho \left[\frac{2-\delta}{2+\delta}\frac{A_0^{1+\frac{\delta}{2}}}{4(4\pi)^{\frac{\delta}{2}} }\right] =: \frac{8\pi \rho}{3} G_\text{eff}.
\end{equation}
In other words, matter field feels an effective gravitational constant\footnote{We remark that Ref.\cite{4}, on the contrary, incorporates the extra term in a new energy density $\rho_\text{DE}$ that is attributed to dark energy, instead of modifying the gravitational constant. We can check that the two are equivalent by taking Eq.(2.12) of \cite{4} and substituting their expressions for $\Lambda$ and $\beta$. The difference between the two expressions is given by the fact that \cite{4} considers only the case $k=0$ and keeps an integration constant $C$ that in \cite{1} is set to $0$ or considered as part of the total energy density $\rho$. In our work, we will use the form of Ref.\cite{1}. As we shall see below, when allowing BEI to run, we would have an extra term that can be interpreted as a dynamical dark energy, different from the identification in \cite{4} for the fixed index case.} $G_\text{eff}$ instead of $G$. 
This is one of the main results of Ref.\cite{1}.

This was obtained by applying the first law of thermodynamics to the energy flux through the apparent horizon of the Friedmann–Lema\^{i}tre–Robertson–Walker (FLRW) universe, instead of the black hole horizon \cite{0807.1232}. The expression for the horizon is
\begin{equation}
    r_\text{AH} = \left(H^2 + \frac{k}{a^2}\right)^{-\frac{1}{2}}. \label{app hor}
\end{equation}
In the following we will assume a spatially flat universe ($k=0$).

Allowing the Barrow entropy index to run, the change in entropy would be:
\begin{flalign}
	dS &= d\left(\frac{4\pi r^2}{A_0}\right)^{1+\frac{\delta}{2}} \\ \notag
	   &= \left(\frac{4\pi r^2 }{A_0}\right)^{1+\frac{\delta}{2}}\left[\frac{(2+\delta)\dot{r}}{r}+\ln\left(\frac{4\pi r^2 }{A_0}\right)\frac{\dot{\delta}}{2}\right]dt. 
\end{flalign}
Then, using the first law of thermodynamics and the continuity equation, we can show, following \cite{1}, that
\begin{flalign}\label{dE}
    dE &= -\frac{1}{2\pi r}\left(\frac{4\pi r^2 }{A_0}\right)^{1+\frac{\delta}{2}}\left[\frac{(2+\delta)\dot{r}}{r}+\ln\left(\frac{4\pi r^2 }{A_0}\right)\frac{\dot{\delta}}{2}\right]dt  \\ \notag
		&= - 4\pi r^3 H(\rho + p)dt,\\ \notag
 \end{flalign}  
and consequently
\begin{equation}
		\frac{(4\pi)^{\frac{\delta}{2}}}{2\pi A_0^{1+\frac{\delta}{2}} }\left[(2+\delta)r^{\delta-3}+\ln{\left(\frac{4\pi r^2}{A_0}\right)} \frac{r^{\delta-2}}{2} \frac{d\delta}{dr}\right]dr= -\frac{\dot{\rho}}{3} dt.
	\end{equation}

We note that $r_\text{AH} = 1/H$. In a sensible cosmology, $H$ decreases with time, so $r_\text{AH}$ is an increasing function of time. In other words, a larger $r$ corresponds to a later time.
From the black hole thermodynamics argument as per Sec.(\ref{2}), we want $\delta$ to at least decrease as $\delta \sim \text{const.}/r^2 \ln(r^2)$ at large $r$. At early time it should approach unity. For definiteness we can take a function that is always smaller than the asymptotic behavior $\text{const.}/r^2 \ln(r^2)$, so a natural choice is $\delta(r)=e^{-r}$. 

The differential form of the modified Friedmann equation is equivalent to
\begin{flalign}\label{dBEI}
	 &(2-\delta)r^{\delta-3} dr + \frac{2-\delta}{2+\delta}\ln\left(\frac{4\pi r^2}{A_0}\right)\frac{r^{\delta-2}}{2} \frac{d\delta}{dr}dr \notag
	\\ &= -\frac{8\pi}{3}\underbrace{\left[\frac{2-\delta}{2+\delta}\frac{A_0^{1+\frac{\delta}{2}} }{4 (4\pi)^{\frac{\delta}{2}}}\right]}_{=:G_\text{eff}} d\rho,
\end{flalign}
where $G_\text{eff}$ is the effective gravitational constant, the same as defined in the fixed BEI case \cite{1}, as in Eq.(\ref{corrected}). We immediately notice that $G_\text{eff}$ is monotonically decreasing as $\delta \to 0$, with $G_\text{eff}(\delta=1)=1/(3\sqrt{\pi})\approx 0.189$ near the Big Bang, which is about 20\% the percent value of unity (in Planck units). We defer the possible cosmological implications of this varying $G_\text{eff}$ to the Discussion section. 

For now we note that since $\delta$ is a function of $H$, the Friedmann equation in the case of a running BEI cannot be easily solved by direct integration. We can, however, still analyze the late time and early time behavior separately.
At late time, we choose some $r^*$ sufficiently large such that $1/r^2 \ln(r^2) \ll 1$ is as small as we wish. 
We have $\delta \sim 0$ (more precisely, we may say that $\delta$ is small enough that it is slowly varying compared to $r$, so we can take some small value of fixed $\delta$ when integrating with respect to $r$).
Then, with $A_0=4$ in Planck units, the first term yields (with $R>r^*$)
\begin{equation}
2\int_{\tilde{r}}^R r^{-3} dr = 2\left[-\frac{1}{2r^2}\right]_{r^*}^R=\frac{1}{{r^*}^2}-\frac{1}{R^2}.
\end{equation}
The r.h.s of Eq.(\ref{dBEI}) integrates to give the matter density 
\begin{equation}
-\frac{8\pi}{3}[\rho]_{{\rho}^*}^{\rho(R)}.
\end{equation}
Thus, in the absence of the second term in Eq.(\ref{dBEI}), this will just give the usual Friedmann equation (as we take $R \to \infty$; note $\rho(R) \to 0$ in this limit because matter density decreases as the universe expands):
\begin{equation}
H^2 = \frac{8\pi \rho}{3}
\end{equation}
evaluated at the time that corresponds to $r^*$.

Therefore, we can interpret the second term in Eq.(\ref{dBEI}) as the differential form of the effective dark energy:
(so that Eq.(\ref{dBEI}) yields the form\footnote{$\frac{1}{3}\int_{\Lambda_\text{eff}(r^*)}^{\Lambda_\text{eff}(R)} d\Lambda_\text{eff} = \frac{1}{3}[{\Lambda_\text{eff}(R)-\Lambda_\text{eff}(r^*)}]$. This means that the observed value of the dark energy would be $\Lambda_\text{eff}(r^*)-\Lambda_\text{eff}(\infty)$ in the limit $R\to \infty$. Without loss of generality we can take $\Lambda_\text{eff}(\infty)\equiv 0$ so that $\Lambda_\text{eff}(r^*)$ is the observed value of the effective cosmological constant.} $H^2-\Lambda/3=8\pi \rho/3$)
\begin{equation}
\frac{d\Lambda_\text{eff}}{3} = \frac{2-\delta}{2+\delta}\ln\left(\frac{4\pi r^2}{A_0}\right)\frac{r^{\delta-2}}{2} \frac{d\delta}{dr}dr.
\end{equation}
If we use the ansatz $1/r^2 \ln(r^2)$ for $\delta$, then
\begin{flalign}
\frac{d\Lambda_\text{eff}}{3} = &-\frac{2-\delta}{2+\delta}\ln\left(\frac{4\pi r^2}{A_0}\right)\\ \notag &\times \frac{r^{\delta-2}}{2} \left[\frac{2}{r^3 \ln(r^2)}+\frac{2}{r^3(\ln(r^2))^2}\right] dr. \notag
\end{flalign}

At late time, $\delta \sim 0$, we have an upper bound:
\begin{flalign}
-\left.\frac{\Lambda_\text{eff}}{3}\right|_{\Lambda(r^*)}^{\Lambda(R)} &\sim \frac{1}{2}\int_{r^*}^R \frac{\ln(\pi r^2)}{r^2}\left[\frac{2}{r^3 \ln(r^2)}+\frac{2}{r^3(\ln(r^2))^2}\right] dr \label{int} \\
&< \frac{1}{2} \int_{r^*}^R  \frac{\ln(\pi r^2)}{r^2} \frac{4}{r^3 \ln(r^2)} dr \\
&= 2 \int_{r^*}^R \frac{1}{r^5}\frac{1}{\ln(r^2)}[\ln \pi + \ln(r^2)] dr\\
&< 2 \int_{r^*}^R \left(\frac{1}{r^5}+\frac{1}{r^5}\right)dr=4 \int_{r^*}^R \frac{1}{r^5} dr\\
&=\left[-\frac{1}{r^4}\right]_{r^*}^R = \frac{1}{{r^*}^4}-\frac{1}{R^4}.
\end{flalign}
That is,
\begin{equation}
\frac{\Lambda_\text{eff}(r^*)}{3} < \frac{1}{r^{*4}}
\end{equation}
as $R \to \infty$. Since $r^*$ is chosen to be very large, $\Lambda_\text{eff}$ must be very small, and furthermore it is getting smaller at later time.
In fact, the integral Eq.(\ref{int}) yields
\begin{equation}
\Lambda_\text{eff}(r^*) \sim 
\frac{3}{4}\left[ \frac{1}{r^{*4}} \frac{\ln(\pi r^{*2})}{\ln(r^{*2})} -2 \left(1-\ln(\pi)\right)
		\text{Ei}\left(-4 \ln(r^*)\right) \right], \notag
\end{equation}
where $\text{Ei}$ is the exponential integral function. One can check numerically that it is positive and decreases towards $0$ very rapidly.
Such an asymptotically vanishing effective cosmological constant scenario was previously discussed in \cite{Wetterich,0911.1063,1704.08040,0707.3148,0401002,frolov}. 
Next we show that although the universe is late time de Sitter (dS)-like, it is anti-de Sitter (AdS)-like in the very early universe. Such a behavior had been observed, for example, in non-commutative quantum field theory \cite{1911.08921}.

In the vicinity of the Big Bang, the effective dark energy term depends on the profile of $\delta$ (how $\delta$ approaches $1/r^2 \ln(r^2)$).
For example, if $\delta=1$ for some time after the Big Bang, then $d\delta/dr = 0$, and there is no effective dark energy.
However, if we choose as we did previously, $\delta=e^{-r}$, then the effective dark energy at the Big Bang is (with $\delta \sim 1$ and $0<\varepsilon \ll 1$)
\begin{equation}
\left.\frac{\Lambda_\text{eff}}{3}\right|_{\Lambda_\text{eff}(\varepsilon)}^{\Lambda_\text{eff}(2\varepsilon)} \sim -\frac{1}{3}\int_\varepsilon^{2\varepsilon} \ln (\pi r^2) \frac{e^{-r}}{2 r} dr. 
\end{equation}
The integration interval is kept small near the Big Bang to ensure that $\delta=e^{-r}$ is nearly constant so that
we can approximate as follows:
\begin{equation}
\left.{\Lambda_\text{eff}}\right|_{\Lambda_\text{eff}(\varepsilon)}^{\Lambda_\text{eff}(2\varepsilon)}\sim -\int_\varepsilon^{2\varepsilon} \ln (\pi r^2) \frac{e^{-r}}{2 r} dr \sim -\varepsilon\ln (\pi \varepsilon^2) \frac{e^{-\varepsilon}}{2 \varepsilon}.
\end{equation}
That is,
\begin{equation}
\frac{\Lambda_\text{eff}(2\varepsilon)-\Lambda_\text{eff}(\varepsilon)}{\varepsilon} \sim -\ln (\pi \varepsilon^2) \frac{e^{-\varepsilon}}{2 \varepsilon}.
\end{equation}
In the limit $\varepsilon \to 0$, this yields $d\Lambda_\text{eff}/d\varepsilon \to \infty$. It follows that\footnote{As per Footnote 4, the observed value of the effective cosmological constant over the interval $(\varepsilon,2\varepsilon)$ is $\Lambda_\text{eff}(\varepsilon)-\Lambda_\text{eff}(2\varepsilon)<0$. Since they are both comparable in magnitude (negative and with the same divergence properties), we can just refer to $\Lambda(\varepsilon)$ as the effective cosmological constant at the Big Bang.} $\Lambda_\text{eff} \to -\infty$ at the Big Bang with a log divergence\footnote{Indeed, 
a function $f(\varepsilon)=\ln(\pi\varepsilon^2)e^{-\varepsilon}<0$ for small $\varepsilon$, satisfies $f(2\varepsilon)-f(\varepsilon) \to \infty$ and $f'(\varepsilon) \to \infty$.
Another example to illustrate the same phenomenon, albeit with a different divergence, is to consider $f(x)=-1/x$. This function satisfies $f(2\varepsilon)-f(\varepsilon) \to \infty$ and $f'(\varepsilon)=1/\varepsilon^2 \to \infty$. Colloquially, $f$ climbs out from $-\infty$ at the origin with an infinite positive slope.}.  It is worth commenting that \cite{2010.00986} gives a strong constraint for the smallness of $\delta$ (assuming $\delta$ to be fixed) during the Big Bang nucleosynthesis (BBN) epoch. Thus, we know that $\delta$ cannot  be slowly varying in the early universe if $\delta$ starts out near unity at the Big Bang. In fact, this is partly the reason why we chose $\delta=e^{-r}$ in our toy model, as it decays rather quickly at the beginning. Here, however, we are using the mathematical fact that any function is nearly constant if the interval of its domain is sufficiently small.

An infinitely large negative cosmological ``constant'' is not good (the universe might immediately re-collapse), though we should also note that this is a logarithmic divergence $\sim O(\ln\varepsilon)$,
which is still ``smaller'' in magnitude than the $1/\varepsilon$-divergence in the modified Hubble term\footnote{Since $H \to \infty$, there is still a Big Bang singularity in this model.} (from the first term in Eq.(\ref{dBEI})), which is proportional to
\begin{equation}
\int_\varepsilon^{2\varepsilon} r^{-2} dr \sim \varepsilon\frac{1}{\varepsilon^2} = \frac{1}{\varepsilon}. 
\end{equation}

Of course, we could in principle choose other functions to obtain a finite nonzero value. Regardless, the integral
\begin{equation}
 \Lambda_\text{eff}(\varepsilon) \sim -\int_\varepsilon^{2\varepsilon} \ln\left({\pi r^2}\right)\frac{1}{2r} \frac{d\delta}{dr}dr.
\end{equation}
is always negative because 
$d\delta/dr < 0$ \emph{and} $\ln(\pi r^2)<0$ for small $r$, as long as $\delta$ is a strictly decreasing function.
Therefore $\Lambda_\text{eff}$ started out negative, crossed zero at some point in time (at $r_H=\sqrt{1/\pi} \approx 0.564$), and at late time it is small and positive. By continuity this suggests that 
$\Lambda_\text{eff}$ reached a maximum value at some point in the past, before decreasing towards its current small value $10^{-122}$. 
Another possibility is that $\delta$ contains a factor that cancels the $\ln(\pi r^2)$ term; then, in principle, the transition for AdS to dS can happen at an even later time depending on the zero of the function. 
Similar scenarios have been proposed in the literature \cite{1105.0078,1105.2636,2001.02451,2008.10832,1912.08751,1309.2732,1808.06623,2112.10641}, with \cite{2001.02451} and \cite{2008.10832} proposing that the transition from AdS phase to dS phase occurred around the recombination epoch, while \cite{1912.08751} argued that the transition occurs as late as at redshift $z\approx 2.32$, which triggered the late-time acceleration. It is also worth mentioning that \cite{2001.02451, 2108.09239} proposed that the AdS-dS transition could help to alleviate the Hubble tension. In fact, \cite{2108.09239} argued that the $S_8$ tension \cite{1409.2769,2005.03751} (equivalently, the $\sigma_8$ tension) is also alleviated, among other improvements; see also \cite{2002.11707}. In \cite{2201.11623}, the authors concluded that solving both the $H_0$ and $\sigma_8$ tensions would require that $G_\text{eff} < G$ at some redshift $z$, if the equation of state satisfies $w(z) \leqslant -1$ (otherwise, $w(z)$ must cross the phantom divide). Thus, it is worth mentioning that a running BEI also causes a varying gravitational constant, and furthermore from Eq.(\ref{corrected}), our $G_\text{eff}$, in the Planck units, always satisfies
\begin{equation}
\frac{1}{3\sqrt{\pi}}<G_\text{eff}=\frac{2-\delta}{2+\delta}\frac{1}{\pi^{\frac{\delta}{2}} } < 1=G.
\end{equation}
for all values of admissible $\delta \in [0,1]$. 
However, there are also arguments which state that this is unlikely to alleviate the Hubble tension by itself \cite{2112.14173}; see also the subtleties discussed in the next section.

\section{Discussion}\label{5}
In this work, we have argued that if the Barrow entropy does encode some form of quantum gravitational correction, then in order to be consistent with black hole thermodynamics of large black holes (for which quantum gravity correction should be small), BEI must vary according to the energy scale. In terms of the horizon area, it should decay at least as fast as $1/(A \ln A)$. 

When applied to the cosmological context, we show that the modified Friedmann equation has many surprising properties: the universe started out with an effective negative cosmological constant, but at late time it becomes positive and small, decaying towards zero asymptotically. As already explored in the literature, such a scenario could help to alleviate the Hubble tension and the $S_8$ tension. See, however, \cite{1907.07953, 2203.10558}.

Despite the fact that BEI is, thus far, only a phenomenological parametrization of quantum gravity effects, it is interesting to see that it could potentially explain some cosmological mysteries. In fact, although the explicit form of $\delta$ can only be determined once we have a better understanding of how to derive the Barrow entropy from a theory of quantum gravity, the features we obtained are quite generic as long as $\delta$ is assumed to be monotonically decreasing with the energy scale. The shortcoming is that we still do not know the exact behavior of $\delta$ especially at early time. Another caveat is that in deriving Eq.(\ref{dE}), we have used the first law as $dE = -T dS = \dot{\rho} V dt$, in which the volume change has been neglected, which amounts to the assumption that $\dot{r} \ll Hr$. This holds at late time with exponential late time expansion, but may not hold near the Big Bang; this will depend on the explicit solution of the scale factor, which in turn depends on the profile of $\delta$. 

Lastly we comment on the varying effective gravitational constant. Whenever there is a varying $G$, one might worry that the continuity equation (the one employed in Eq.(\ref{dE}) is the standard one $\dot{\rho} = -3H(\rho + p)$ that follows from $\nabla_{\mu}T^{\mu\nu} = 0$) may be modified since $\nabla_\mu(G T^{\mu\nu})$ may no longer imply $\nabla_\mu(T^{\mu\nu})=0.$ However the ``bare'' gravitational constant $G$ that appears at the level of the action is not the same as the effective $G_\text{eff}$ that appears in the modified Friedmann equation Eq.(\ref{dBEI}). Also, the Barrow entropy itself contains $A_0=4G$, which is the ``bare'' $G$. The fact that the $G_\text{eff}$ is smaller in the early universe could suggest that we may be able to ameliorate the arrow of time problem \cite{feynman,Price1, Price2, vaas, Price3, wcc1, 0611088, 123, 0711.1656, 1406.3057} following the same line of thought of Greene et. al. \cite{0911.0693} as well as Sloan and Ellis \cite{1810.06522}. (Note that to ameliorate the $H_0$ tension, one usually requires that $G_\text{eff}<G$ at late time. However, in our case, this condition always holds, even in the early universe.) However, this would require a very careful examination of the structure formation process by studying the perturbation equation. Typically, the effective gravitational constant that governs structure formation may not be the same as the $G_\text{eff}$ that appears in the background Friedmann equation, which may also differ from the ``bare'' $G$. A concrete example is provided in \cite{2202.04908}. Structure formation in the case of fixed BEI has been recently studied in \cite{2205.04138}. In fact, a varying gravitational constant would cause many issues and early time cosmology would need to be re-examined in close details \cite{1009.5514} (for example, CMB angular power
spectrum would be modified \cite{1009.5514,2112.14173}).

In conclusion, the effects of running BEI requires a deeper investigation. There is also a need to better understand how Barrow entropy can possibly arise from a theory of quantum gravity. Does the fractalized geometry only hold for horizons or any spacetime hypersurface in general? What about spacetime manifold itself? Barrow entropy is not the first proposal that involves fractal geometry in gravity. It has previously been proposed that spacetime dimension becomes fractalized and decreases towards the Planck scale \cite{0811.1396,1009.1136,1505.05087,1602.01470,1702.08145,1904.04379}, which is a different modification (a 2-sphere in such a fractalized geometry has a \emph{lower} dimension \cite{0811.1396}, not higher as per Barrow's proposal), but perhaps one could obtain some ideas about how to derive Barrow entropy from these different approaches. Connections with fractional quantum mechanics \cite{2107.04789} and other notions of entropies in relation to holographic dark energy \cite{2112.10159,2105.08438,2205.08876,2201.02424} should also be further investigated.

\begin{acknowledgments}
		YCO thanks the National Natural Science Foundation of China (No.11922508) for funding support. He also thanks Brett McInnes for useful discussions.
	\end{acknowledgments}

\end{document}